# Magnetostrictive Phononic Frequency Combs

Guanqi Ye, Ruitong Sun, Junning Zhao, and Fusheng Ma[*]

*Key Laboratory of State Manipulation and Advanced Materials in Provincial Universities, Institute of Physics Frontiers and Interdisciplinary Sciences, School of Physics and Technology, Nanjing Normal University, Nanjing 210046, China*

*Email: phymafs@njnu.edu.cn

**ABSTRACT:**

Magnetostriction, mechanical-to-magnetic or magnetic-to-mechanical response, plays a pivotal role in magneto-mechanical systems. Here, we propose and experimentally demonstrate a magneto-mechanical frequency comb via the three-wave mixing mechanism, which solely requires the involvement of the fundamental mode $f_0$ of a magnetostrictive macroresonator. Two types of combs, *i.e.*, the integer-harmonic combs and the half-integer-harmonic combs, are observed in kHz regime with Hz resolution by magnetically pumping the mm-scale resonator with near-resonant $f_p \approx f_0$ and modulating $f_s \ll f_0$. The integer-harmonic combs are centered at $lf_p$ ($l = 1, 2, 3, …$), while the half-integer-harmonic combs are centered at $(2n - 1) f_p/2$ ($n = 1, 2, 3, …$) resulting from the period-doubling bifurcation of $f_p$. The tooth spacing of both types of combs is determined and can be continuously tuned by changing $f_s$ from Hz to kHz. Moreover, the half-integer-harmonic combs can be purposely switched with frequency shifting half a tooth spacing via suppressing period-doubling bifurcation. The experimentally observed formation, evolution, and switching of combs can be well understood by introducing the bias magnetical force and modulated linear stiffness into the Duffing equation. Our findings on magnetically manipulated phononic frequency comb could provide a magneto-mechanical platform for potential non-invasive and contactless sensing and even antenna for wireless operation.

## Introduction

Frequency comb is a spectrum composed of equidistant narrow lines[1]. Optical frequency combs (OFCs) have been widely utilized for frequency metrology, optical clocks, spectroscopy, optical ranging, astronomical spectrograph calibration, and optical frequency synthesis[2-7]. Inspired by the achievements of OFCs, phononic frequency combs (PFCs)[8,9], also called acoustic or mechanical frequency combs, are attracting increasing attention and have been realized in elecro- and opto-mechanical systems. For elecro-mechanical systems, a mechanical mixer has been reported using a strongly driven nanomechanical resonator of silicon-on-insulator wafers[10]. Based on the theory of Fermi-Pasta-Ulam $\alpha$ chains, the PFCs were theoretically proposed[11] and experimentally demonstrated using a piezo-electrically driven micromechanical resonator[12]. Since then, PFCs have been widely studied in electro-mechanical systems on manometer and micrometer scale via $N$-wave mixing, bifurcation, internal resonance, and parametric resonance[12-22]. PFCs have also been realized in nonlinear opto-mechanical systems through the Kerr nonlinear effect[23,24] and the mechanical mode lasing[8,25-29] resulting from the parametric coupling between phonons and photons. Recently, benefiting from the advantages of non-invasive and contactless manipulation, magneto-mechanical systems are emerging as a platform for mechanical applications as magnetoelectric sensors, logic-in-memory devices, and magnetometers[30-34]. Therefore, PFCs based on magneto-mechanical systems are expected but not yet approached.

Most of the reported PFCs are working in the megahertz[8,12,20] and gigahertz[25] range. Actually, the Hz to kilohertz (kHz) range is crucial to many applications of magnetic field detection[35], geological surveying[36], magnetoencephalography[37], and so forth. Specially, PFCs working in kHz are highly anticipated in underwater ranging and communication due to the weak attenuation and the low dispersion in liquid[38,39]. Based on the magneto-mechanical coupling, the mechanical resonators made of magnetostrictive materials have been utilized in wireless neural stimulation[40,41] and antenna[42] in the frequency range from Hz to kHz with scale from centimeter to millimeter. The magnetization induced by mechanical deformation can be wirelessly detected by near-field inductive coupling through a coil with high precision. Therefore, magnetostrictive materials are promising for investigating PFCs in kHz range without requiring high-cost manufacturing and measuring technologies.

From the application point of view, PFCs with precisely controlled tooth spacing are expected to enhance the precision in sensing and metrology[39]. Additionally, PFCs with the capability to switch between different comb modes can further enrich the range of applications. For the reported mechanical system based on mode coupling and competition with single driving, the comb tooth spacing is hard to freely manipulate since it is determined by the discrete eigenfrequency[17,25,26]. Recently, it has been demonstrated that the comb tooth spacing of PFCs can be tuned continuously by changing modulating frequency in electro-mechanical systems with multiple driving[16,43]. Whereas, PFCs with simultaneous tunability of both comb center and comb tooth spacing remain limited in single-mode mechanical systems.

In this work, we propose a scenario of switchable PFCs only involving the fundamental mechanical mode $f_0$ of the resonator, which is experimentally demonstrated in the kHz regime using a magneto-mechanical macrosystem on millimeter scale. Two types of PFCs resulting from three-wave mixing and period-doubling (PD) bifurcation are observed, *i.e.*, integer-harmonic combs (IHCs) and half-integer-harmonic combs (HIHCs) by magnetically pumping a magnetostrictive resonator with two-tone near-resonant (NR) $f_p \approx f_0$ and modulating $f_s \ll f_0$. The tooth spacing of both types of combs is determined by $f_s$, while the centers of IHCs and HIHCs are located at $lf_p$ and $(2n - 1)f_p/2$ ($l, n$ = 1, 2, 3, …), respectively. Furthermore, the HIHCs can be purposely switched with frequency shifting half a tooth spacing by suppressing PD bifurcation. By modifying the Duffing equation through introducing the bias magnetical force and modulated linear stiffness, the formation, evolution, and switching of the observed PFCs are qualitatively explained. The magnetically manipulated mechanical frequency combs observed here could pave the way for non-contacting sensing and metrology.

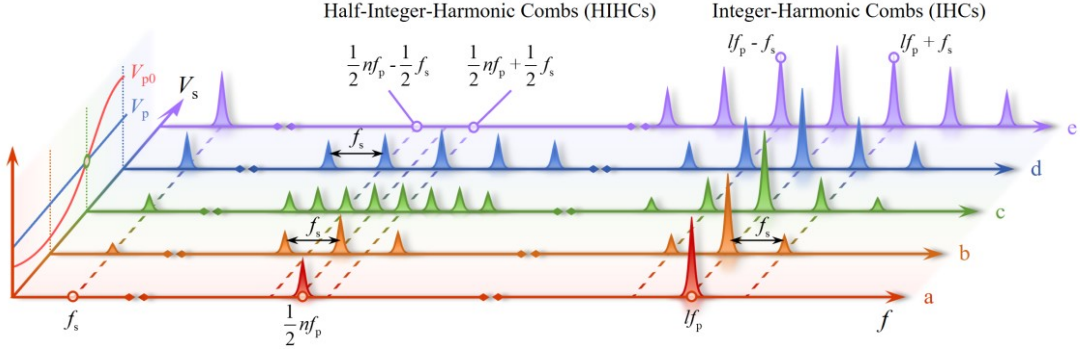

**Fig. 1 | Concept of switchable PFCs via PD bifurcation.** Two types of PFCs with tooth spacing $f_s$: the HIHCs $(2n - 1)f_s/2 \pm mf_s$ and $(2n - 1)f_s/2 + f_s/2 \pm mf_s$; the IHCs $lf_p \pm mf_s$.

# Results

## Scenario of generating PFCs

Considering a mechanical resonator with the fundamental mode $f_0$, it would be driven into the nonlinear dynamic regime and generate PFCs by two-tone pumping fields, *i.e.*, a NR pumping $V_{np} = V_p\cos(2\pi f_p t)$ and a modulating pumping $V_{mp} = V_s\cos(2\pi f_s t)$, as schematically shown in Fig. 1. Here, the two pumping frequencies satisfy the condition of $f_p \approx f_0$ and $f_s \ll f_0$. Keeping the NR pumping amplitude $V_p$ as a constant, there will be a threshold $V_{p0}$ corresponding to the PD bifurcation point and varying with the modulating pumping amplitude $V_s$. For $V_s = 0$ and $V_p < V_{p0}$, the forced vibration mode and its higher harmonics $lf_p$ ($l = 1, 2, 3, …$) are excited. For $V_s = 0$ and $V_p > V_{p0}$, the resonator will transit to the PD bifurcation state, which is manifested by the appearance of the extra PD vibration modes $(2n - 1)f_p/2$ ($n = 1, 2, 3, …$) as shown in Fig. 1a. For $V_s > 0$ and $V_p > V_{p0}$, both IHCs and HIHCs are generated via three-wave mixing as shown in Fig. 1b. The IHCs and HIHCs can be described in the manner of $lf_p \pm mf_s$ and $(2n - 1)f_p/2 \pm mf_s$, respectively, where $m = 1, 2, 3, …$ represents the $m^{th}$ comb tooth. The bandwidth of IHCs grows steadily with increasing $V_s$ as shown in Figs. 1b-e. In contrast, the HIHCs will switch between distinct states besides the increasing of bandwidth with $V_s$. Firstly, the HIHCs switch from $(2n - 1)f_s/2 \pm mf_s$ to $(2n - 1)f_s/2 + f_s/2 \pm mf_s$ when $V_{p0}$ crosses $V_p$ as shown in Figs. 1b and d. To be noted that the switching can be approached by either superposition or chaos for $V_{p0}$ around $V_p$. As an instance, Fig. 1c exhibits the superposition of $(2n - 1)f_s/2 \pm mf_s$ and $(2n - 1)f_s/2 + f_s/2 \pm mf_s$.

Secondly, the HIHCs will be gradually suppressed and finally disappear when $V_s$ further increases as shown in Fig. 1e. For the sake of simplicity, Fig. 1 only shows the $l = 1$ and $n = 1$ case.

**Magneto-mechanical system**

In order to demonstrate the proposed scenario of generating switchable PFCs, we build up a magneto-mechanical system consisting of an amorphous magnetostrictive ribbon and three coils as schematically shown in Fig. 2a. Here, we use the AYFY-N Metglas with dimensions of 15 mm × 4 mm × 25 μm as the magneto-mechanical resonator. Two transmission (Tx) coils are utilized to excite the Metglas: $H_p = h_p\cos(2\pi f_p t)$ as Tx1 for NR pumping; and $H_s = h_s\cos(2\pi f_s t)$ as Tx2 for modulating pumping, respectively. The pumping magnetic fields $H_p$ and $H_s$ are proportional to the applied voltage $V_p$ and $V_s$, respectively, while the bias magnetic field $H_{dc}$ is provided by an electromagnet. Based on electromagnetic induction and inverse magnetostriction effect, the deformation-induced magnetization at frequency $f$ is detected and converted to electrical signal $V_{in}(f)$ by the receiving (Rx) coil.

The fundamental mechanical mode $f_0$ of the used resonator is determined by Young's modulus, which can be modulated by the bias magnetic field $H_{dc}$. The measured dependences of $f_0$ and $df_0/dH_{dc}$ on $H_{dc}$ are shown in Fig. 2b, where the $f_0$ is extracted from the $S_{11}$ spectra. The left inset in Fig. 2b presents the mode profile of $f_0$ simulated by using the finite element method[44]. The right inset in Fig. 2b shows the measured $S_{11}$ coefficient spectrum with $H_{dc} = 14.3$ Oe, at which the sensitivity of the resonator to the external magnetic field is the highest. To make sure the sensitivity of the resonator, the $H_{dc}$ used in this work is fixed at 14.3 Oe. It can be fitted out from the spectrum that $f_0$ is 150.4 kHz with a mechanical quality factor of 109. To be noted that the $f_0$ of the resonator can be manipulated from MHz to Hz by scaling up the resonator from micrometers to meters (see Fig. S1 in Supplemental Material).

**PD bifurcation**

Firstly, we study the effect of $V_p$ on the vibration state of the resonator with $f_p = 150$ kHz and $V_s = 0$. Fig. 2c shows the color plot of measured $V_{in}(f)$ spectra as a function of $V_p$, which can be divided into three regimes. For $V_p < 9.5$ V, the resonator is forced into the periodic vibration state with $f = f_p = 150$ kHz. For $9.5$ V $< V_p < 21.1$ V, the $V_{in}(f)$ spectra exhibit two new tones $f = 0.5f_p = 75$ kHz and $f = 1.5f_p = 225$ kHz, which indicates the appearance of the PD bifurcation of $f_p$. Higher harmonics are also observed in the high-frequency range. For $V_p > 21.2$ V, the $V_{in}(f)$ spectrum displays typical noise, *i.e.*, broadly distributed spectrum, which indicates that the vibration bifurcates on the period-doubling route to chaos. The three distinct states of period, PD bifurcation, and chaos are clearly shown in Fig. 2d with $V_p = 5$, 18, and 25 V, respectively.

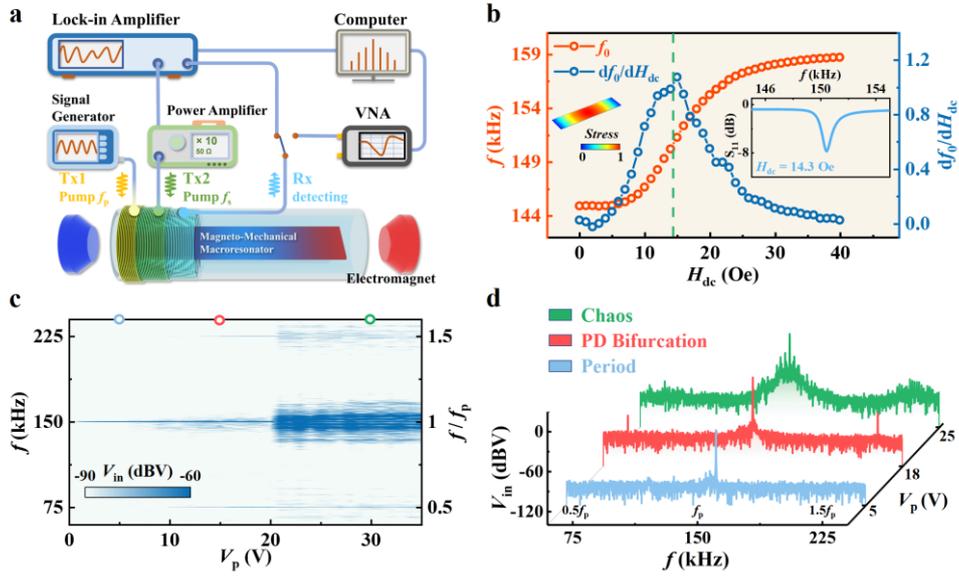

**Fig. 2 | Period-doubling bifurcation induced by single near-resonant pumping. a** Schematic of the experimental setup for PFCs using magneto-mechanical macroresonator. Electromagnet provides bias magnetic field $H_{dc}$; signal generator and lock-in amplifier provide periodic pumping magnetic fields. **b** Measured dependence of $f_0$ and $df_0/dH_{dc}$ on $H_{dc}$. Right inset displays a typical $S_{11}$ spectrum measured at $H_{dc} = 14.3$ Oe; left inset displays the simulated mode profile of $f_0$ with normalized stress magnitude shown in color. **c** Color plot of measured $V_{in}(f)$ as a function of $V_p$ for $f_p = 150$ kHz and $H_{dc} = 14.3$ Oe. **d** Three typical $V_{in}(f)$ spectra extracted from **c** for $V_p = 5$, 18, and 25 V corresponding to different vibration states: period, PD bifurcation, and chaos, respectively.

**Formation and evolution of IHCs and HIHCs**

Secondly, we further study the effect of $V_p$, $V_s$, and $f_p$ on the PD bifurcation states of the resonator. Fig. 3a shows the measured $V_{in}(f)$ spectra for various $V_s$ with $f_p = 150$ kHz, $V_p = 20$ V, and $f_s = 200$ Hz. It is observed that the $l = 1$ IHCs $f_p \pm mf_s$ are generated and the comb bandwidth grows steadily by increasing $V_s$. For $V_s = 750$ mV, the bandwidth can approach 20 kHz with the number of comb teeth exceeding 100 (see the details as S3 in Supplemental Material). Broadband PFCs could be realized by further increasing $V_s$. In contrast to the PFCs based on mode coupling whose tooth spacing is determined by the discrete eigenfrequency of the resonator, the tooth spacing in our work can be continuously tuned by changing $f_s$.

Interestingly, it is observed that the HIHCs multiply switches between mode M1 ($f_p/2 \pm mf_s$) and mode M2 ($f_p/2 + f_s/2 \pm mf_s$) by increasing $V_s$. As shown in Fig. 3a, for $f_s = 200$ Hz, the mode M1 centered at 75 kHz is observed for $V_s = 150$ and 300 mV, while the mode M2 centered at 75.1 kHz is observed for $V_s = 240$ mV. During the transition between modes M1 and M2, there is also a mixed mode ($f_p/2 \pm mf_s/2$) as the superposition of modes M1 and M2 for $V_s = 260$ mV as well as a chaotic mode for $V_s = 230$ mV. The evolution process of HIHCs can be clearly observed by color plotting the measured $V_{in}(f)$ spectra in the form of $\Delta = f - f_p/2$ as a function of $V_s$ for $f_p = 150$ kHz and $V_p = 10$ V as shown in Fig. 3b. The HIHCs switch between modes M1 and M2 four times when $V_s$ increases from 0 to 580 mV, and finally disappear for $V_s > 580$ mV. A typical chaotic transition between modes M1 and M2 is exhibited as inset in Fig. 3b. We also extract the amplitudes of three comb teeth ($\Delta = -0.1, 0,$ and $0.1$ kHz) as shown in Fig. 3c. The amplitude of $\Delta = 0$ tooth abruptly jumps five times when $V_s$ increases from 0 to 580 mV. While the amplitude of $\Delta = -0.1$ and $0.1$ kHz teeth undergo four times jumping. For each jump, the PD bifurcation point $V_{p0}$ across the $V_p$. For $V_s > 580$ mV, the amplitude sum of $\Delta = -0.1$ and $0$ kHz teeth decreases to zero indicating that the HIHCs disappear. The asymmetry of HIHCs, the larger number and amplitude of comb teeth on the high-frequency side are attributed to a red-shift detuning $f_p < f_0$[43]. The evolution of HIHCs is also highly dependent on $V_p$. When $V_p$ increases from 10 to 20 V, the number of comb teeth increases and the HIHCs switch 7 times as shown in Figs. 3d and e. To study the effect of detuning $f_p - f_0$ on HIHCs, we also measured $V_{in}(f)$ spectra as a function of $f_p$ with a step size of 0.1 kHz for $V_p = 10$ V and $V_s = 200$ mV as shown in Fig. 3f. The HIHCs are generated in the range from $f_p = 148.5$ to $154.6$ kHz. This is further confirmed by the variation amplitude of the comb teeth $\Delta = -0.1, 0,$ and 0.1 kHz as shown in Fig. 4g. It should be noted that the tooth spacing of HIHCs can be continuously tuned by changing $f_s$ (see Fig. S4 in Supplemental Material).

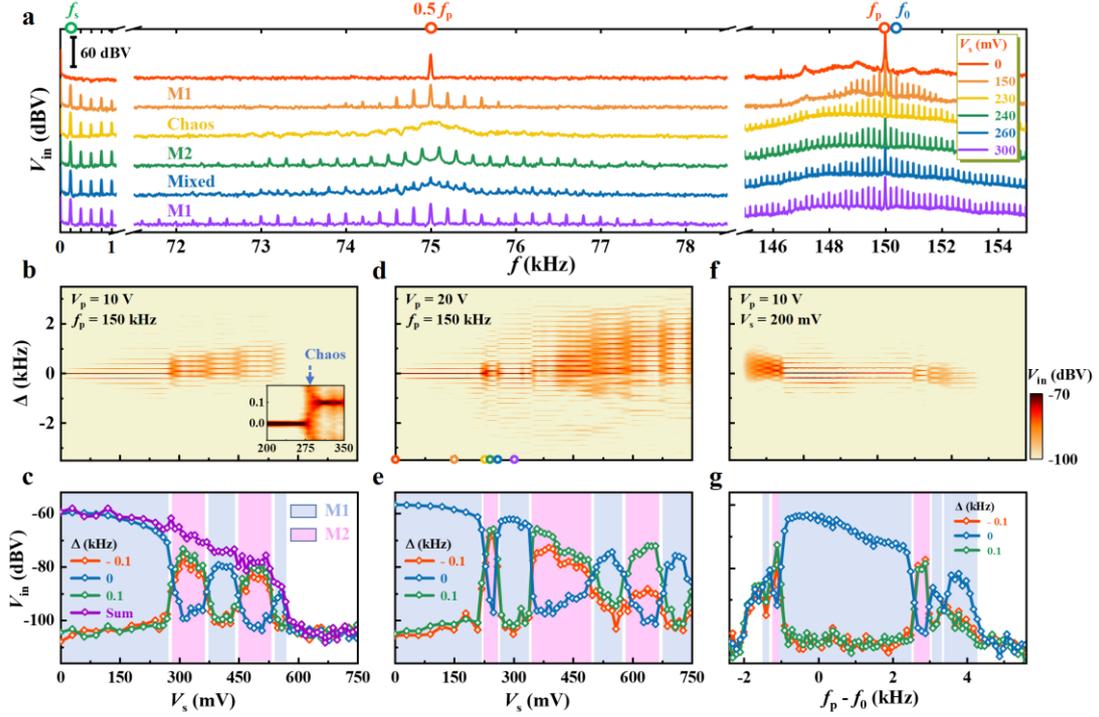

**Fig. 3 | IHCs and HIHCs realized via three-wave mixing. a** $V_{in}(f)$ spectra under various $V_s$ for $f_p$ = 150 kHz and $V_p$ = 20 V. HIHCs show two distinct modes labeled by M1 and M2 as well as two transient modes labeled by mixed and chaos. Each spectrum shifted 60 dBV apart for clear comparison. Color plot of $V_{in}(f)$ spectra in the form of $\Delta = f - f_p/2$ as a function of $V_s$ for $f_p$ = 150 kHz, **b** $V_p$ = 10 V, and **d** $V_p$ = 20 V, respectively. Inset in **b** shows a zoom-in view of the chaotic transient in HIHCs. Color circles in **d** correspond to the spectra in **a**. Variation of teeth amplitude as a function of $V_s$ for the three teeth: **c/e** $\Delta$ = - 0.1, 0, and 0.1 kHz in **b/d**. Color plot of $V_{in}(f)$ spectra in the form of $\Delta = f - f_p/2$ as a function of $f_p$ for $V_p$ = 10 V and $V_s$ = 200 mV. **g** Variation of teeth amplitude as a function of $f_p$ for the three teeth $\Delta$ = - 0.1, 0, and 0.1 kHz in **f**. Blue and pink background represent the regions of modes M1 and M2, respectively. $f_s$ is fixed at 200 Hz. All spectra are averaged 4 times to reduce noise.

**Theory based on modified Duffing equation**

Nonlinearities in mechanical systems are associated with the Duffing mechanism as a consequence of lattice anharmonicity, *i.e.*, a nonlinear term in Hook's law[45]. The nonlinear dynamics of mechanical resonators can be modeled with a generalized Duffing equation

$$\frac{d^2x}{dt^2} + \gamma \frac{dx}{dt} + (2\pi f_0)^2 x + \beta x^3 = A_p \cos(2\pi f_p t), \qquad (1)$$

where $x$ is displacement, $t$ is time, $\gamma$ is damping, $\beta$ is nonlinear coefficient, $f_0$ is eigenfrequency of mechanical mode, $A_p$ is driving amplitude, and $f_p$ is driving frequency. For our magneto-mechanical system, we introduce the bias magnetic force $A_0$ and the modulated linear stiffness $A_s\cos(2\pi f_s t)$ into Eq. (1) and obtain a modulated Duffing equation as

$$\frac{d^2x}{dt^2}+\gamma\frac{dx}{dt}+(2\pi f_0)^2 x+\beta x^3+A_s x\cos(2\pi f_s t) = A_p\cos(2\pi f_p t)+A_0, \qquad (2)$$

We carry out the calculation by solving Eq. (2) using fourth order Runge-Kutta method, and the used parameters are: $\gamma = 7 \times 10^3$, $f_0 = 150.4$ kHz, $f_p = 150$ kHz, $\beta = 4 \times 10^{22}$, $f_s = 200$ Hz, and $A_0 = 1 \times 10^7$.

To understand the experimentally observed HIHCs, we calculate the vibration spectra, temporal waveforms, and orbital phase portraits of four different dynamical states, *i.e.*, period, PD bifurcation, comb mode M1, and comb mode M2, as shown in Fig. 4. For $A_s = 0$ and $A_p = 1.8 \times 10^7$ as shown in Fig. 4a, there is no vibration at $f_p/2 = 75$ kHz in the spectrum, which is confirmed by the presence of only $T_p = 1/f_p$ and $T_p/2 = 1/(2f_p)$ in the temporal waveform corresponding to the limit cycle and the small circular orbit in the phase portrait, respectively. For $A_s = 0$ and $A_p = 1.9 \times 10^7$ as shown in Fig. 4b, the vibration spectrum exhibits a tone at $f_p/2$ corresponding to the period-doubling of $T_p$ and a bifurcation of the limit cycle. For $A_s = 8 \times 10^{11}$ and $A_p = 1.9 \times 10^7$ as shown in Fig. 4c, the comb mode M1 ($f_p/2 \pm mf_s$) exhibits in the spectrum. There is a pulse train with a pulse period of $T_s = 1/f_s$ in the temporal waveform, and the bifurcated limit cycle in the phase portrait is contracted. For $A_s = 9 \times 10^{11}$ and $A_p = 1.9 \times 10^7$ as shown in Fig. 4d, the comb mode M2 exhibits in the spectrum. The modes M1 and M2 cannot be distinguished in the temporal waveform due to an offset of only 100 Hz in the vibration spectra. Whereas, the bifurcated limit cycles of mode M2 are merged to one due to the shift of the PD bifurcation point. Similar to the experimental observations, the calculated evolution of HIHCs switching between modes M1 and M2 is also found as a function of $A_s$ (see calculated results as S5 in Supplemental Material). It should be noted that the calculation parameters of $\gamma$, $\beta$, $A_p$, $A_s$, and $A_0$ are chosen to qualitatively reproduce the experimental results. In addition, the experimentally observed IHCs can also be explained by solving Eq. (2) without considering the $\beta x^3$ term (see the details as S6 in Supplemental Material).

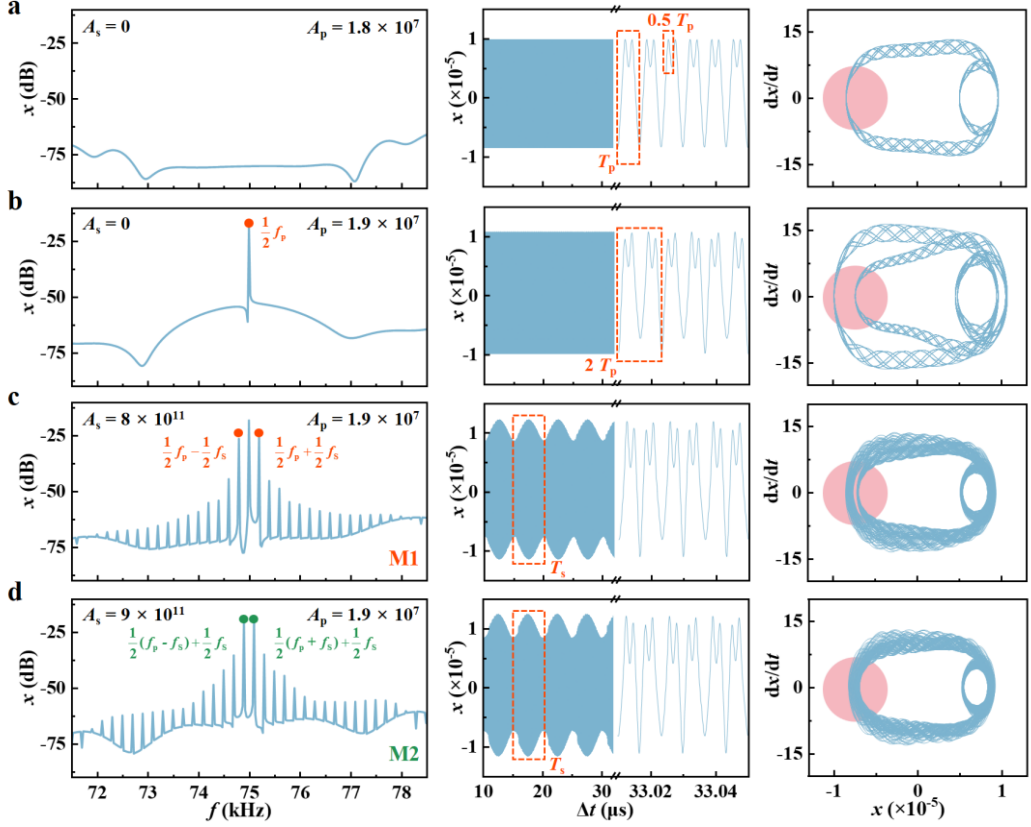

**Fig. 4 | Calculation with modified Duffing equation.** Left, middle, and right columns present the calculated vibration spectra, temporal waveforms, and orbital phase portraits, respectively. The calculated evolution of HIHCs: **a** Periodic state, **b** PD bifurcation, **c** Comb mode M1, **d** Comb mode M2. Red circles highlight the differences between the four portraits.

## Conclusion

In summary, two types of FPCs based on three-wave mixing and PD bifurcation are proposed and experimentally demonstrated, *i.e.*, IHCs and HIHCs. We build up a magneto-mechanical macrosystem consisting of a 15 mm magnetostrictive resonator with fundamental mechanical mode 150.4 kHz and two magnetical pumpings, *i.e.*, NR pumping $f_P$ and modulating pumping $f_s$. The bandwidth of IHCs $lf_P + mf_s$ can approach 20 kHz with the number of comb teeth ~ 100. Interestingly, the HIHCs can switch between modes $(2n-1)f_s/2 \pm mf_s$ and $(2n-1)f_s/2 + f_s/2 \pm mf_s$ resulting from the shift of bifurcation point. The formation, evolution, and switching of the observed PFCs can be qualitatively explained by the modified Duffing equation including the bias magnetical force and the modulated linear stiffness. This work contributes significantly to the field of PFCs utilizing magneto-mechanical macrosystems, and the reported nonlinear mechanism of switching PFCs could be extended to other mechanisms involving either single or multiple eigenfrequencies of resonator.

## Methods

The magneto-mechanical macroresonator used in this study is an amorphous magnetostrictive ribbon (Metglas AYFA-N) fabricated by pulse laser. The used dimension is: 15 mm × 4 mm × 25 μm. Fig. 2a shows the schematic of the experimental setup for the generation and measurement of the PFCs. Three coils are arranged on the outside of the Metglas along the length. In detail, the Tx1 coil is utilized to apply NR pumping $H_p = h_p\cos(2\pi f_p t)$ via a function waveform generator (DG5000, RIGOL). The Tx2 coil is utilized to apply modulating pumping $H_s = h_s\cos(2\pi f_s t)$ via lock-in amplifier (UHF 600MHz, Zurich) and wide band amplifier (ATA-1372A, Aigtek). The pumping magnetic fields $H_p$ and $H_s$ are proportional to the applied voltage $V_p$ and $V_s$, respectively. Based on electromagnetic induction and inverse magnetostriction effect, the deformation-induced magnetization at frequency $f$ is detected and converted to electrical signal $V_{in}(f)$ by the Rx coil, then input into the VNA or lock-in amplifier. Besides, the bias magnetic field $H_{dc}$ is provided by an electromagnet (P9060, East Changing). The measured response measured by the lock-in amplifier is post-processed, using a LabView software interface, in the time and frequency domains to obtain the FFT and frequency response of the mechanical resonator.

## Data availability

All relevant data are available from the corresponding authors on request.

## Acknowledgements

This work was supported by the National Key Research and Development Program of China (Grant No.2023YFF0718400) and the National Natural Science Foundation of China (NNSFC) (Grant Nos. 12474119 and 12074189).

## Author contributions

G. Ye and F. Ma conceived and designed the research. G. Ye performed the experimental measurements and theoretical calculations. G. Ye, R. Sun, J. Zhao and F. Ma performed the analysis of results and the writing of the manuscript. All authors have read and approved the final manuscript.

## Competing interests

The authors declare no competing interests.